# Power Law Liquid – A Unified Form of Low-Energy Nodal Electronic Interactions in Hole Doped Cuprate Superconductors

T.J. Reber[1], X. Zhou[1], N.C. Plumb[1,&], S. Parham[1],J.A. Waugh[1], Y. Cao[1], Z. Sun[1,#], H. Li[1], Q. Wang[1], J.S. Wen[2], Z.J. Xu[2], G. Gu[2], Y. Yoshida[3], H. Eisaki[3], G.B. Arnold[1], D. S. Dessau[1]*

1 Dept. of Physics, University of Colorado, Boulder, 80309-0390, USA
2 Condensed Matter Physics and Materials Science Department, Brookhaven National Labs, Upton, New York, 11973 USA
3 AIST Tsukuba Central 2, 1-1-1 Umezono, Tsukuba, Ibaraki 3058568, Japan
& Now at Swiss Light Source, Paul Scherrer Institut, CH-5232 Villigen PSI, Switzerland
# Now at University of Science and Technology of China, Hefei, China.

* Correspondence to: dessau@colorado.edu

The strange-metal phase of the cuprate high temperature superconductors, above where the superconductivity sets in as a function of temperature, is widely considered more exotic and mysterious than the superconductivity itself. Here, based upon detailed angle resolved photoemission spectroscopy measurements of $Bi_2Sr_2CaCu_2O_{8+\delta}$ over a wide range of doping levels, we present a new unifying phenomenology for the non-Fermi liquid normal-state interactions (scattering rates) in the nodal direction. This new phenomenology has a continuously varying power law exponent (hence named a Power Law Liquid or PLL), which with doping varies smoothly from a quadratic Fermi Liquid to a linear Marginal Fermi Liquid and beyond. Using the extracted PLL parameters we can calculate the optics and resistivity over a wide range of doping and normal-state temperature values, with the results closely matching the experimental curves. This agreement includes the presence of the T* "pseudogap" temperature scale observed in the resistivity curves including the apparent quantum critical point. This breaks the direct link to the pseudogapping of antinodal spectral weight observed at similar (but here argued to be different) temperature scales, and also gives a new direction for searches of the microscopic mechanism at the heart of these and perhaps many other non-Fermi-liquid systems.

**Introduction:**

The non-superconducting state of the cuprates is debatably more intriguing and unusual than the high temperature superconducting state itself. One of the most iconic aspects, and a key to the electronic interactions out of which the superconducting state is borne, is the non-Fermi liquid resistivity above $T_c$ [1,2]. This non-Fermi Liquid behavior, often called the "strange metal" state, is characterized by a linearly increasing resistivity with temperature, counter to the quadratic behavior expected for electron-electron scattering in Landau's Fermi liquid (FL) theory[3]. This linear-in-T "strange metal" behavior is considered so unusual that it is believed by many to signal a new state of matter, motivating many of the most influential and exotic theoretical ideas of the cuprate problem including Anderson's resonating valence bond (RVB) [4], Varma's Marginal Fermi liquid (MFL) [5], many ideas about quantum critical points [6,7] as well as duals of string-theory models of quantum gravity [8]. Linearity in the measured resistivity has also recently been found in numerous other compounds, with connections to the cuprate physics sought [9].

The strange metal state with linear-in-T scattering occurs near the middle of the doping phase diagram, roughly above where the optimal or maximum $T_c$ exists. To the far right at high doping levels a regular (quadratic in T) Fermi liquid exists, while to the left at low doping levels is an unusual and poorly understood pseudogap state in which there is an incomplete suppression of low-energy spectral weight, especially at the "antinodal" regime of the Brillouin zone. And while tremendous effort has been invested to understand the pseudogap and strange metal states [10,11], a great amount of confusion still exists as even evidenced by a lack of agreement about the doping phase diagram of these effects, especially whether these states intersect the superconducting dome or whether they continue beyond it (see supplemental materials section S2 for a discussion of this).

Here we utilize angle resolved photoemission spectroscopy (ARPES) to study the electronic scattering rates or "self energies" of the $Bi_2Sr_2CaCu_2O_8$ family of cuprate superconductors. The unique momentum-selectivity of ARPES allows it to measure the scattering rates in a simple and direct manner simply by looking at peak widths. We took advantage of the special ability of ARPES to measure the scattering rates as a function of both energy and temperature, whereas previous studies have relied on either the temperature or energy dependences alone.

**Electronic interactions from ARPES:**

Fig 1a presents ARPES data from an optimally doped $Bi_2Sr_2CaCu_2O_{8+\delta}$ sample taken in the normal state at T=100K. The data were taken along the nodal direction where the d-wave superconducting gap is zero. We use low energy (7 eV) photons, which give enhanced energy resolution, momentum resolution, and bulk sensitivity compared to regular ARPES [12]. A slice through this spectrum at constant energy, known as a momentum distribution curve (MDC) is generally a Lorentzian whose width is $\Gamma_{MDC}$, with this width directly proportional (through the electron velocity dE/dk – see supplemental materials section S3) to the single particle scattering rate, or equivalently, the imaginary part of the electron self energy Σ"(ω).

Fig 1b present a compilation of the energy and temperature dependence of Σ"(ω) from four differently doped samples. As a function of energy each spectrum shows an approximately linear behavior at high energy with an upward curvature of the scattering rates near $E_F$ that are reminiscent of the Fermi liquid $\omega^2$ dependence. Thus this data is consistent with our intuition that the low energy states retain some Fermi-liquid-like features, while the higher energy ones show the non-Fermi-liquid linear dependence. As the temperature of any one sample is increased the curves shift up to higher scattering rates. Such a full set of ARPES scattering rate data as a function of energy and temperature (and doping) has not, to our knowledge, been previously presented.

**Form of Electronic Interactions:** Taking inspiration from (a) the success of the phenomenological "Marginal Fermi Liquid" form of self-energy proposed by Varma [5], and (b) the non-integral power laws in Anderson's "Hidden Fermi Liquid" [13], we propose the following "Power Law Liquid" or PLL form for the electronic scattering rates:

$$\Sigma''_{PLL}(\omega) = \Gamma_0 + \lambda \frac{\left[(\hbar\omega)^2 + (\beta k_B T)^2\right]^\alpha}{(\hbar\omega_N)^{2\alpha-1}} = \Gamma_0 + \lambda \frac{\left[(k_B T)^2 \left[\left(\frac{\hbar\omega}{k_B T}\right)^2 + \beta^2\right]\right]^\alpha}{(\hbar\omega_N)^{2\alpha-1}} \tag{1}$$

where Σ"(ω) is the imaginary part of the self energy directly measured in our experiment, $\Gamma_0$ is an offset parameter accounting for impurity or disorder scattering, λ is a coupling parameter indicating the overall strength of the scattering, $\omega_N$ is a normalization frequency whose exponent maintains the proper dimensionality of the self-energy, parameter β governs the comparative strengths of temperature and energy, and α is the critical PLL variable that takes the system from a FL to a MFL, and beyond. Note that this formalism does not contain any low energy scales that would be associated with superconductivity, the pseudogap, phonons, or other bosonic modes. The only energy scale is $\omega_N$ which is not strictly necessary, but we include to maintain constant units for coupling parameter λ as a function of doping. Of all the energy scales in the system the one that closest matches our $\omega_N$ is the full conduction bandwidth relative to $E_F$. This form exhibits ω/T scaling only - a phenomenon commonly encountered in quantum-critical types of theories.

The black dotted lines of fig 1b show fits of the data to the PLL self-energy, which reproduce the data extremely well up to the rather large scale of 0.1 eV scale that is considered here (even Fermi Liquid theory is typically only considered to be relevant to roughly 10 meV). This large energy range does pass the scale of potential bosonic modes (phonons, magnetic resonance), and while these have a noticeable effect on the spectra deep in the superconducting state they couple only weakly at these higher normal-state temperatures, where the scattering is strongly dominated by the electronic interactions that are discussed here.

All curves for one sample have been fit simultaneously to equation 1, greatly constraining the parameter set that can fit the data. The extracted parameters as a function of doping are shown in figure 1c. One set of parameters $\alpha$,$\beta$,$\lambda$, and $\Gamma_0$ is obtained per sample, with these four parameters fitting all the ARPES data for all energies and all normal state temperatures. We do not include the normalization frequency $\omega_N$ here because it has been fixed at the high energy of 0.5 eV for all samples, which is purposely far beyond the 0.1 eV energy scale over which the data are fit, minimizing its impact on the obtained physics. $\omega_N$ is also fully mathematically irrelevant for the case that the parameter $\alpha$=1/2 and almost irrelevant when $\alpha$ is near 1/2, as is the case for most important doping values.

Figure 1c shows that $\lambda$, and $\beta$ are essentially independent of doping level (and also energy and temperature). The $\beta$ values (3.5±0.5) are close to the theoretical expectation of $\pi$ for a Fermi liquid metal (blue dashed line in fig 1c), with this result based upon the conversion of the Matsubara frequencies from the imaginary to real axes [14]. The main experiments that have addressed this issue in the literature are optics experiments, and then even for the simple Fermi Liquid case these have not successfully found the expected scaling between T and $\omega$ [see ref 15 and our supplemental materials section S4]. Therefore, the $\beta$ values uncovered here serve as a combination of a classic theoretical prediction, constraints on new theories, and a confirmation of the reasonableness of the PLL form of interactions. The offset parameter $\Gamma_0$ ranges from about 8 to 35 meV as a function of doping, as shown by the offset lines in panel b and discussed more extensively in the supplemental materials section S5.

Parameter $\lambda$ is essentially constant as a function of doping, pegged to the value 0.5, again confirming the simplicity and universality of the form of interactions. This value is roughly consistent with the recent notions of "Planckian dissipation" that stated that the optimally doped cuprates had the maximum possible scattering rate, determined by Planck's constant [16]. The fact that $\lambda$ is constant across the phase diagram implies that all doping levels may have this same maximum electronic coupling, though the form of the coupling (controlled by $\alpha$) varies strongly as a function of doping.

The parameter $\alpha$ is the most important. As shown in figure 1c, $\alpha$ takes on a roughly linear dependence that increases with doping, and is very close to 1/2 at optimal doping. In this case eqn 1 reduces to the hyperbolic form:

$$\Sigma''_{Opt} = \Gamma_0 + \lambda\sqrt{(\hbar\omega)^2 + (\beta k_B T)^2} \qquad (2)$$

which is *linear* in both energy (for T=0) and temperature (for ω=0), i.e. it is of the MFL type of interaction [5] (see fig 2b) and the parameter $\omega_N$ also becomes irrelevant. If we extrapolate α to very high doping levels such that $\alpha$=1, the PLL form becomes:

$$\Sigma''_{HOD} = \Gamma_0 + \lambda \frac{(\hbar\omega)^2+(\beta k_B T)^2}{\omega_N} \qquad (3)$$

and the quadratic dependence on energy and temperature of the normal Fermi liquid is recovered (fig 2a). As we go to the underdoped region the self-energy takes on an unusual S-curve shape with energy (fig 2c) that has the expected sub-linear behavior at higher energies. This behavior is a natural aspect of the PLL self-energy and is not possible with a linear combination of quadratic Fermi Liquid and linear MFL. To emphasize the S-curve in the underdoped regime we draw in the linear extrapolation of the deep energy dependence in Fig. 2c at a finite temperature of 100 K. Surprisingly, this form still has a quadratic-like behavior at low energies as shown in figure 2c, reminiscent of (but different than) a Fermi Liquid.

The data shown in figure 1b is from 27 individual "cuts" of data, each of which contains on order of 50 energy points and 100 k points, or over $10^5$ data points total. Characterizing all of this data with the 5 parameters of equation 1 is impressive. That four of them essentially drop out leaving just the one linearly varying parameter α is even more so.

**Temperature dependence and comparison to resistivity.** Fig 3a shows the calculated normal state temperature dependence of the ω=0 self-energy, using α values according to the linear fit (red dashed line) of fig 1c. We also fixed $\lambda$ to 0.5 and $\beta$ to π, and ignored the impurity scattering term $\Gamma_0$ because, as discussed in the supplemental materials section S5, this term is mostly from forward scattering contributions and/or chemical potential inhomogeneity that have minimal effects on the measured resistivity. Therefore there is essentially only one parameter $\alpha$ that created the entire set of curves shown in fig 3a. These curves have a roughly linear form at higher temperatures, with a deviation from linearity at lower temperature. The approximate temperature at which this deviation occurs is indicated by the arrows (see supplementary materials section S7), although this is not a sharply defined temperature scale.

With this information about the self-energy it is possible, within a few simple models, to calculate the temperature-dependent electrical resistivity. We do this in two standard but simplistic ways, shown for the Boltzmann transport model in fig 3b and the Drude model in fig 3c. The difference in these models comes largely from the way variations in the properties around the Fermi surface are considered, as discussed in the supplemental materials section S6.

Figure 3d presents the measured resistivity of similar samples [17]. The overall scale and shape of the measured resistivity is surprisingly similar to that calculated from the self-energies (figs 3b and 3c). To our knowledge this is the closest agreement yet between the results of a transport measurement and a high-energy spectroscopy such as ARPES, also putting strong constraints on the origin of the "strange metal" resistive fluctuations. This agreement indicates that the electrical resistivity of the cuprates can be closely connected to the single-particle electronic relaxation rates, with these relaxation channels dominated by large-angle scattering (since forward scattering contributes very weakly to the resistivity – see supplemental materials section S5). This would appear to make it difficult for theories with dominant coupling to q~0 fluctuations [18,19,20] to connect to our data.

Conventionally, the temperature scale at which the resistivity deviates from the high temperature linear regime has been noted as T* and has been one of the major tools used to determine the onset of the pseudogap phase. Using the same methods to extract this scale as used in transport, (supplemental materials section S7) a similar temperature scale vs. doping can be extracted from the PLL self-energies, as shown in fig 3a and summarized in Fig. 3e, but called T' here for clarity. Note here that that we have followed these temperature scales through the superconducting dome as if the superconductivity didn't disrupt the PLL phenomenology. Also plotted in fig 3e is an extracted temperature T" in the overdoped regime, where the curvature turns upwards at low temperatures and hence also deviates from linearity. While the experimental transport data of figure 3d don't go far enough into the overdoped regime for the T'' scale to become visible, overdoped data from other families of the cuprates clearly show this behavior [21].

The extracted temperatures shown in fig 3e produce a v-shaped structure, with the crossover between the two branches (where there is perfect linearity) reaching T=0 near optimal doping. This v-shaped fan extending to zero temperature is reminiscent of the quantum critical behavior that has been extensively discussed in the cuprates [6]. These T' and T" values as a function of doping look like the T* from other spectroscopies (fig S1b). However, in contrast to the nodal scattering rate measurements discussed here, measurements of the spectral weight pseudogap from the gapping of energy spectra in the antinodal regime (also from ARPES on $Bi_2Sr_2CaCu_2O_8$) find a T* that asymptotically approaches the superconducting dome, such as shown in fig S1a [11]. While the two different views of the phase diagram could previously have been justified as coming from different samples or types of spectroscopies, we now see that ARPES on BSCCO can give both types of phase diagrams, and that they therefore are both correct but are just measuring different phenomena. This therefore breaks the link between the temperature scales observed in transport (via scattering rates) and the temperature onset of a spectral weight pseudogap in underdoped samples (both of which have historically been called T*).

**Quasiparticle residue and comparison to optics.**

With the normal state nodal spectral function as written down in the PLL form we can determine many properties in addition to resistivity, with two of them described here. The quasiparticle residue Z is a concept from Landau's Fermi liquid theory telling the quasiparticle weight, which in a true Fermi Liquid must be finite but potentially very small (like in a heavy fermion). Again assuming we can kill the superconductivity so as to stay in the PLL phase all the way to zero temperature, we calculate Z (supplemental materials section S8) with results shown in Fig 4a. The PLL Z is finite for all overdoped samples and is identically zero (a true non-Fermi liquid) for all underdoped samples. The transition between them occurs at $\alpha$=1/2 and T=0 and could be considered a quantum critical point, possibly connecting to the v-shaped form of fig 3e that is also reminiscent of quantum criticality.

Despite the unusual power laws in the scattering rates, the PLL spectral weight shows a regular (non-power-law) metallic Fermi edge in all cases (see supplemental information section S9), consistent with the ARPES data. This is however different from other models in which power law scattering rates have been proposed, such as the Projected Fermi Liquid of Anderson [22] and the famous Luttinger Liquid that is known to exist in 1D [23].

Using PLL self energies, we can also simulate the optical conductivity (supplemental materials section S10) as a function of doping, as shown in fig 4b for a simulation temperature T=300K, compared to measured optical data (fig 4c) at the same temperature/dopings [24]. The overall agreement is highly satisfactory, including both the magnitude of the conductivity and the width of the low energy peaks. In general there has been a great deal of debate about the nature of the low energy spectral peaks, especially whether these should be considered Drude peaks representative of a quasiparticle-based Fermi Liquid. Here we see that even though in the strictest sense there are no quasiparticle peaks for $\alpha$<0.5 (i.e. the zero temperature quasiparticle residues vanish), there are zero-frequency upturns in the optical simulations and data that look something like a Drude peak.

**Possible origins of PLL phenomenology**

Just as the origin of the MFL phenomenology is still not known, we expect the origin of the PLL phenomenology will take time to sort out. However, we hope that the extra constraints of this new phenomenology compared to that of the MFL phenomenology will give a boost to our theoretical understanding of these materials.

One precedent for a self-energy with a varying power-law exponent is Luttinger Liquid (LL) physics with spin-charge separation [23], though this is only believed to exist in 1-dimension and so is not directly applicable. However, keeping certain aspects of the LL ideas and extending them to the higher dimensional world of the cuprates is a natural path. Simulations of the doped Hubbard model show similarities to one-dimensional classical spin models and hence pick up the power-law self-energies for the conduction electrons [25]. Another possible path is that of the anti–de-Sitter/conformal field theory (AdS/CFT) correspondence, in which varying power laws of the type observed here would correspond to scanning over a family of possible theories with a gravity dual [8].

A widely discussed explanation for the linear MFL behavior near optimal doping is based upon fluctuations above a quantum critical point, in which two phases meet at zero temperature from opposite sides of the phase diagram [6]. The v-shaped quantum-critical-like behavior shown in fig 3e as well as in fig S1b seem to support this possibility, though here we note that our results are inconsistent with the underdoped regime being Fermi-Liquid-like. Rather, our results indicate that huge doping regimes are z=0 non-Fermi-Liquids and so we may be better off discussing these materials in the context of a set of quantum critical phases [25,26] rather than as a single quantum critical point.

**Outlook.**

While the PLL ansatz is not necessarily a unique choice for a self-energy, it has a remarkable simplicity that with one smoothly varying parameter as a function of doping (the power law exponent α), we can understand the salient normal-state features of the strange-metal ARPES, transport, and optics data. Understanding the origin of this power law behavior therefore holds great promise for understanding the cuprates – not just for the strange metal normal state behavior but also for the superconducting state since it is born out of this power-law strange-metal state.

---

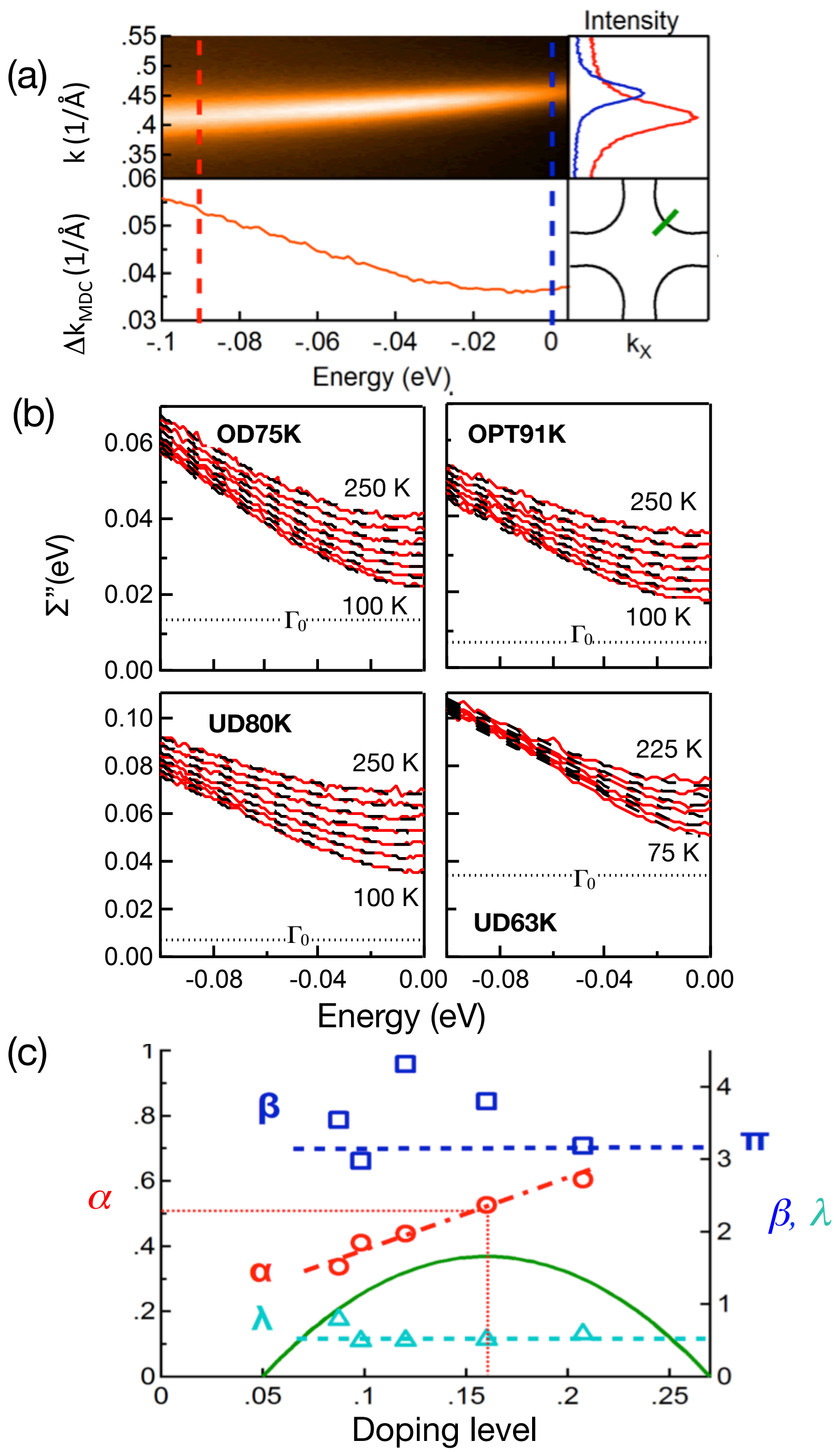

**Fig 1. Experimental electronic self energy Σ" as a function of energy, temperature, and doping.** a) The self energy is extracted from an ARPES spectrum (top left) located at the node (green cut, bottom right) by taking momentum cuts at constant energy (top right) to extract momentum widths $\Gamma_{MDC}(\omega)$ (bottom left) which are directly proportional to Σ"(ω). (b) Measured temperature and energy dependence of Σ" for four different samples from overdoped $T_c$=75K (OD75K) through optimal doped $T_c$=91K (OPT91K) to underdoped $T_c$=63K (UD63K). (c) Fit results for the three main parameters in the model as a function of doping. The superconducting dome is schematically illustrated by the inverted parabola. The most relevant parameter is the power α which is seen to have a simple linear dependence on doping with value 0.5 very near optimal doping.

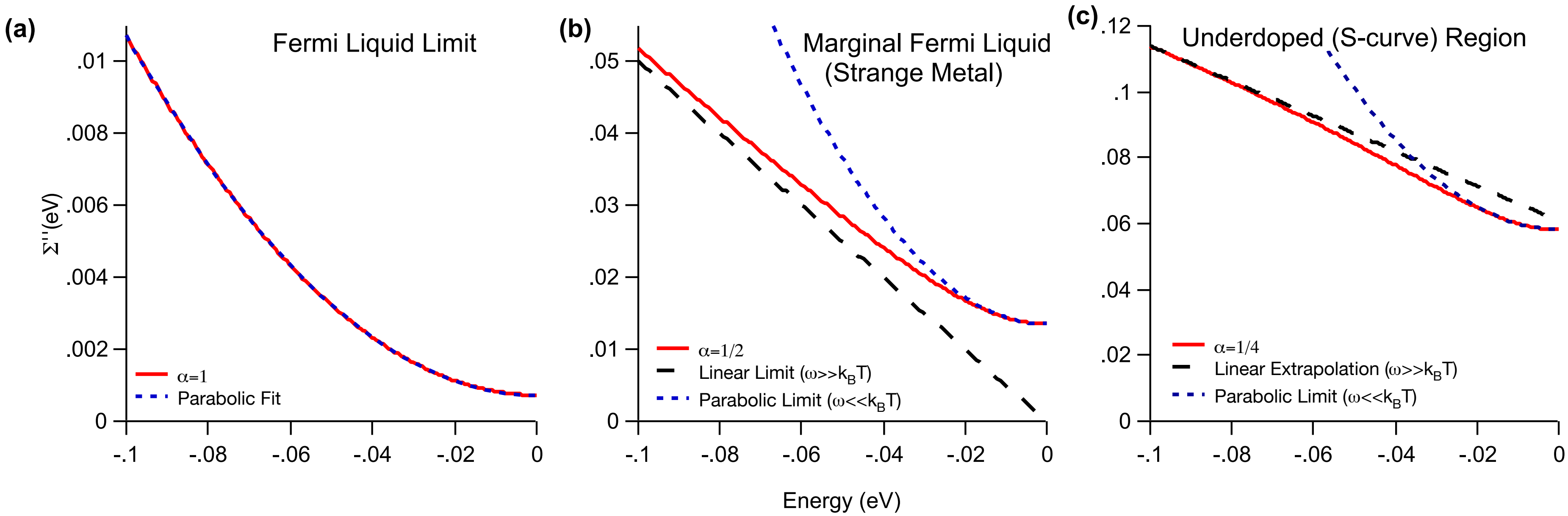

**Fig 2. Form of the self energy for different doping regimes** (red curves) for $T$=100K, $\beta = \pi$, $\lambda = 0.5$ and $\Gamma_0$=0. (a) In the extreme heavily OD case ($\alpha$=1) the system behaves as a normal Fermi liquid with a $\omega^2$ dependence. (b) In the optimal case ($\alpha$=0.5) the system can be approximated as having a $\omega^2$ dependence at low $\omega$ and linear at high $\omega$. (c) In the underdoped case ($\alpha$=0.25) the system has an unusual s-curve $\omega$ dependence shows a concave-up $\omega^2$ dependence at low energies and a concave-down form at higher energy.;

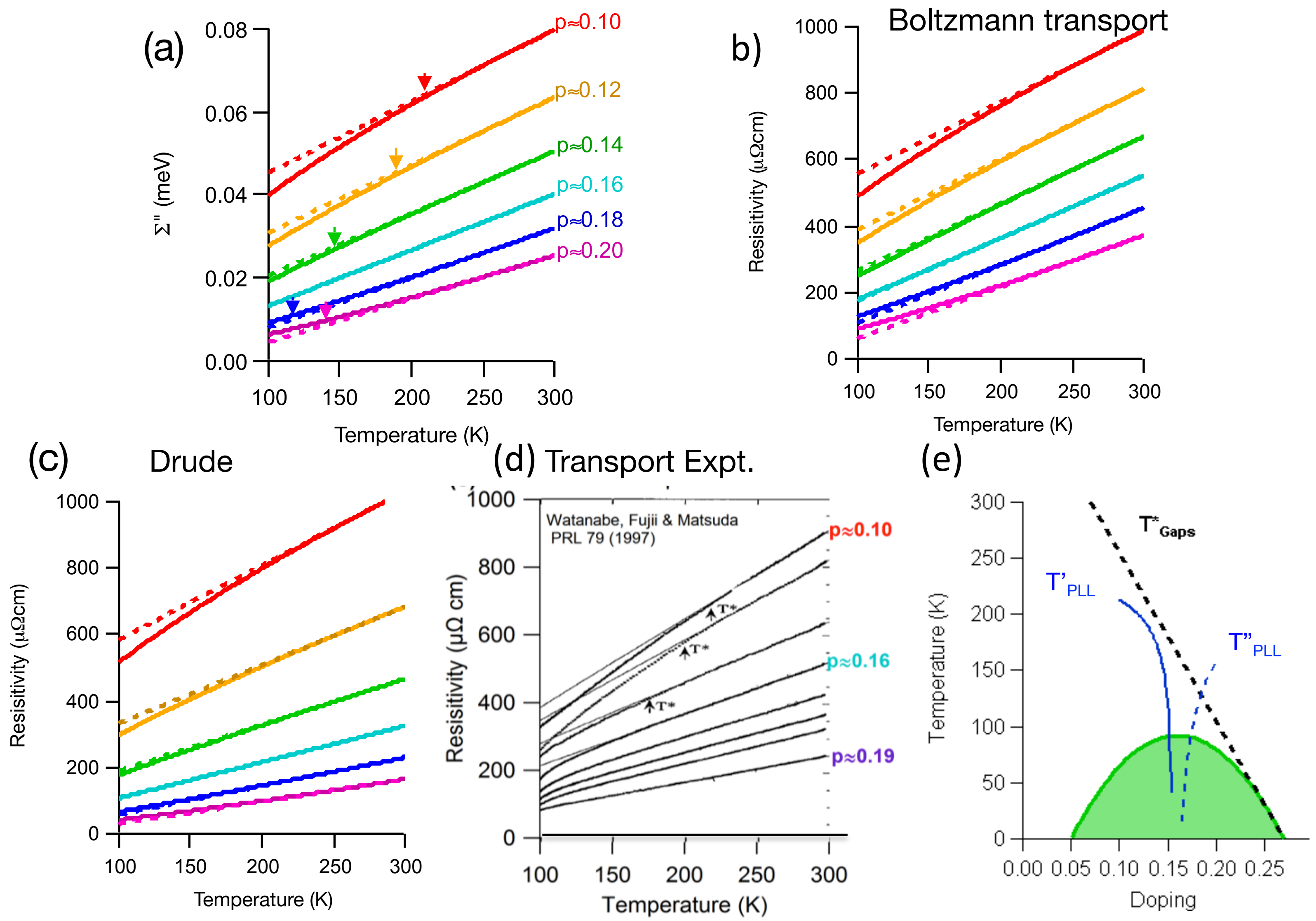

**Fig 3. Scattering rates, resistivity, and apparent temperature scales.** a) Calculated doping and temperature dependence of the nodal Σ"(ω) (solid line) in the ω=0 limit using the linear relation between $\alpha$ and doping level p. Linear fits between [250K, 300K] region are extrapolated as dash lines. Additionally, $\Gamma_0$=0, β=π, and λ=0.5 for all curves. b, c) Resistivity as a function of temperature and doping calculated using two different methods (solid lines, see text for details). Linear extrapolations from [250K, 300K] are shown as dash lines. The temperature dependence is dominated by the temperature dependence of Σ". (d) Resistivity measurements from Watanabe et al. as well as the "pseudogap" temperature scale T* [28]. (e) Compilation of the "break" temperatures from panel a: T' and T" are temperatures where there is an apparent break in Σ" from the more linear form that is observed at high temperatures (up to 300K).

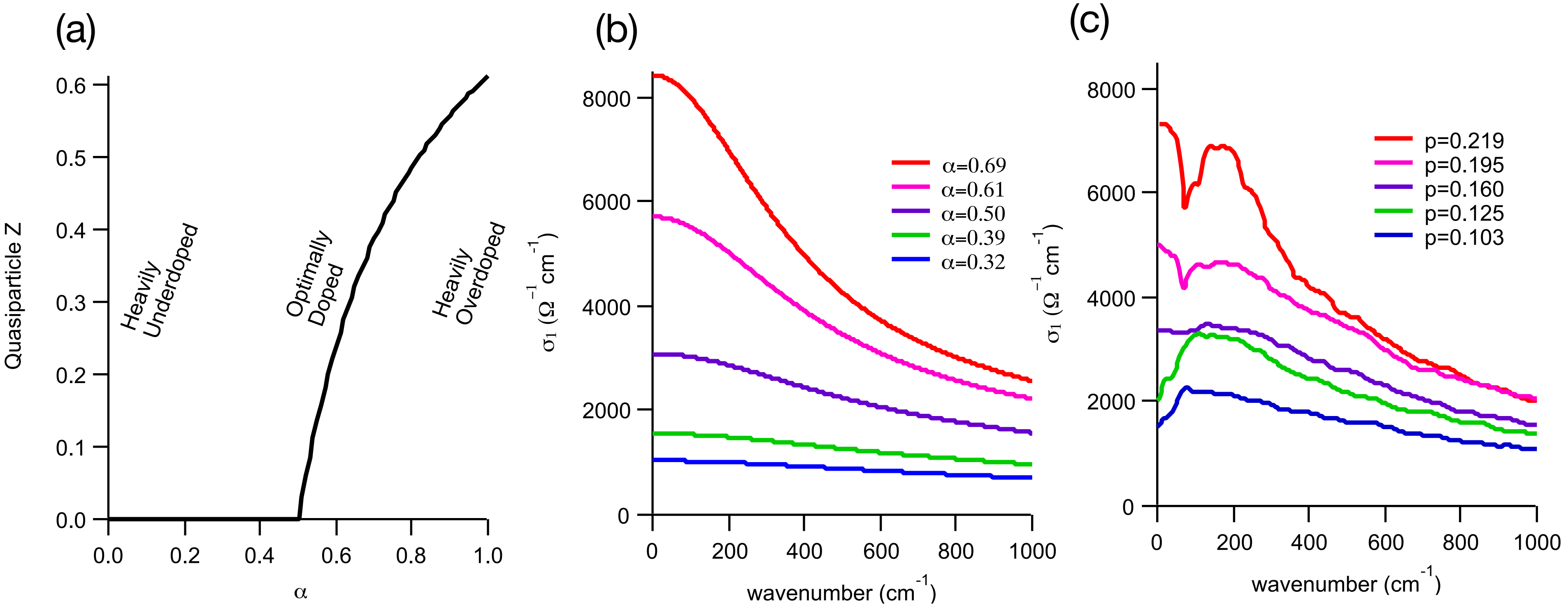

**Fig 4.** a) The quasiparticle residue Z calculated using our PLL functional form (also see supplemental information section S10). b) Simulated and c) measured normal state (T=300K) optical conductivity (reconstructed from ref.[24], see supplementary information section S10) as a function of doping. In panel b) $\alpha$ values were chosen to match the reported doping levels in panel c).